\documentclass[12pt,prd,superscriptaddress,nofootinbib,tightenlines,floatfix,preprintnumbers]{revtex4}

\usepackage{amstext,amssymb}
\usepackage{amsmath}
\usepackage{graphicx}
\usepackage{dcolumn}
\usepackage[hyperfootnotes=false]{hyperref}
\usepackage{xspace}
\usepackage{braket}
\usepackage{color}
\usepackage{slashed} 

\usepackage{slashed} 
\newcommand{\Lmu}{L_\mu - L_\tau}
\begin{document}
\title{Gauged $U(1)_{L_\mu - L_\tau}$ model in light of muon $g-2$ anomaly, neutrino mass and dark matter phenomenology}
\author{Sudhanwa \surname{Patra}}
\email{sudha.astro@@gmail.com}
\affiliation{Center of Excellence in Theoretical and Mathematical Sciences,
Siksha 'O' Anusandhan University, Bhubaneswar-751030, Odisha, India}

\author{Soumya \surname{Rao}}
\email{soumya.rao@adelaide.edu.au}
\affiliation{ARC Centre of Excellence for Particle Physics at the Terascale, Department of Physics, \\
University of Adelaide, Adelaide, SA 5005, Australia}

\author{Nirakar \surname{Sahoo}}
\email{ph13p1005@iith.ac.in}
\affiliation{Department of Physics, Indian Institute of Technology Hyderabad, Kandi, Sangareddy, 502285, Medak, Telengana, India}

\author{Narendra \surname{Sahu}}
\email{nsahu@iith.ac.in}
\affiliation{Department of Physics, Indian Institute of Technology Hyderabad, Kandi, Sangareddy, 502285, Medak, Telengana, India}

\begin{abstract}
Gauged $U(1)_{L_\mu - L_\tau}$ model has been advocated for a long time in light of muon
$g-2$ anomaly, which is a more than $3\sigma$ discrepancy between the experimental
measurement and the standard model prediction. We augment this model with three
right-handed neutrinos $(N_e, N_\mu, N_\tau)$ and a vector-like singlet fermion $(\chi)$
to explain simultaneously the non-zero neutrino mass and dark matter content of the
Universe, while satisfying anomalous muon $g-2$ constraints. It is shown that in a large
parameter space of this model we can explain positron excess, observed at PAMELA,
Fermi-LAT and AMS-02, through dark matter annihilation, while satisfying the relic density
and direct detection constraints.   
\end{abstract}

\maketitle
\newpage
\tableofcontents
\newpage

\section{Introduction}\label{sec:introduction}
The standard model (SM) of elementary particle physics, which is based on the gauge group
$SU(3)_C\times SU(2)_L \times U(1)_Y$ is very successful in explaining the fundamental
interactions of nature. With the recent discovery of Higgs at LHC, the SM seems to be
complete. However, it has certain limitations. For example, the muon $g-2$ anomaly, which
is a discrepancy between the observation and SM measurement with more than $3 \sigma$
confidence level~\cite{Miller:2007kk}. Similarly, it does not explain sub-eV masses of
active neutrinos as confirmed by long baseline oscillation
experiments~\cite{Fukuda:2001nk}. Moreover, it does not accommodate any particle candidate
of dark matter (DM) whose existence is strongly supported by galaxy rotation curve,
gravitational lensing and large scale structure of the universe~\citep{DM_review}. In
fact, the DM constitutes about $26.8\%$ of the total energy budget of the universe as
precisely measured by the satellite experiments WMAP~\cite{wmap} and PLANCK~\cite{PLANCK}. 

At present LHC is the main energy frontier and is trying to probe many aspects of physics
beyond the SM. An attractive way of probing new physics is to search for a $Z'$-gauge
boson which will indicate an existence of $U(1)$ symmetry. Within the SM, we have
accidental global symmetries $U(1)_B$, where $B$ is the baryon number, and $U(1)_L$, where
$L=L_e+ L_\mu+L_\tau$ is the total lepton number. Note that $U(1)_B$ and $U(1)_L$ are
anomalous and can not be gauged without adding any ad hoc fermions to the SM.  However, the
differences between any two lepton flavours, i.e., $L_i- L_j$, with $i,j=e,\mu,\tau$, are
anomaly free and can be gauged without any addition of extra fermions to the SM. Among
these extensions the most discussed one is the gauged $L_\mu -
L_\tau$~\cite{Baek:2001kca,Ma:2001md,Heeck:2011wj,Heeck:2010ub, Heeck:2010pg,Ota:2006xr,Rodejohann:2005ru,Xing:2015fdg,Rivera-Agudelo:2016kjj,Choubey:2004hn,He:1991qd,Araki:2015mya,Fuyuto:2015gmk,Heeck:2016xkh,Altmannshofer:2014cfa, Crivellin:2015lwa,Crivellin:2015mga,Heeck:2014qea,Shuve:2014doa,Salvioni:2009jp,Yin:2009mc,Bi:2009uj,Adhikary:2006rf,Ma:2001tb,Bell:2000vh,He:1994aq,Kim:2015fpa,Harigaya:2013twa,Fuki:2006xw,Elahi:2015vzh,Altmannshofer:2016oaq}
The interactions of corresponding gauge boson $Z'$ are restricted to only $\mu$ and $\tau$
families of leptons and therefore it significantly contribute to muon $g-2$ anomaly, which
is a discrepancy between the observation and SM measurement with more than $3 \sigma$
confidence level. Moreover, $Z'$ does not have any coupling with the electron family.
Therefore, it can easily avoid the LEP bound: $M_Z'/g' > 6$ TeV~\cite{Carena:2004xs}. So,
in this scenario a $Z'$- mass can vary from a few MeV to TeV which can in principle be
probed at LHC and at future energy frontiers.

In this paper we revisit the gauged $U(1)_{L_\mu - L_\tau}$ model in light of muon $g-2$
anomaly, neutrino mass and DM phenomenology.  We augment the SM by including three right
handed neutrinos: $N_e$, $N_\mu$ and $N_\tau$, which are singlets under the SM gauge
group, and a vector like colorless neutral fermion $\chi$. We also add an extra SM singlet
scalar $S$. All these particles except $N_e$, are charged under $U(1)_{L_\mu - L_\tau}$, though singlet
under the SM gauge group. When $S$ acquires a vacuum expectation value (vev), the
$U(1)_{L_\mu - L_\tau}$ breaks to a remnant $Z_2$ symmetry under which $\chi$ is odd while
all other particles are even. As a result $\chi$ serves as a candidate of DM. The
smallness of neutrino mass is also explained in a type-I see-saw framework with the
presence of right handed neutrinos $N_e$, $N_\mu$ and $N_\tau$ whose masses are generated
from the vev of scalar field $S$. 

In this model the relic abundance of DM ($\chi$) is obtained via its annihilation to muon
and tauon family of leptons through the exchange of $U(1)_{L_\mu - L_\tau}$ gauge boson
$Z'$. We show that the relic density crucially depends on $U(1)_{L_\mu - L_\tau}$ gauge
boson mass $M_Z'$ and its coupling $g'$. In particular, we find that the observed relic
density requires $g' \gtrsim 5 \times 10^{-3}$ for $M_Z' \gtrsim 100$ MeV. However, if $g'
\lesssim 5 \times 10^{-3}$ then we get an over abundance of DM, while these couplings are
compatible with the observed muon $g-2$ anomaly. We resolve this conflict by adding an
extra singlet scalar $\eta$ doubly charged under $U(1)_{L_\mu - L_\tau}$, which can drain
out the large DM abundance via the annihilation process: $\overline{\chi}\chi \to
\eta^\dagger \eta$. As a result, the parameter space of the model satisfying muon $g-2$
anomaly can be reconciled with the observed relic abundance of DM. We further show that
the acceptable region of parameter space for observed relic density and muon $g-2$ anomaly
is strongly constrained by null detection of DM at Xenon-100~\cite{Aprile:2012nq} and
LUX~\cite{Akerib:2013tjd}.  Moreover, the compatibility of the present framework with
indirect detection signals of DM is also checked. In particular, we confront the
acceptable parameter space with the latest positron data from
PAMELA~\cite{Adriani:2008zr,Adriani:2010ib}, Fermi-LAT~\cite{FermiLAT:2011ab} and
AMS-02~\cite{Aguilar:2013qda,Accardo:2014lma}.

The paper is arranged as follows. In section-II, we describe in details the different
aspects of the model. Section-III is devoted to show the allowed parameter space from muon
$g-2$ anomaly. In section-IV, we estimate the neutrino mass within the allowed parameter
space. Section V, VI and VII are devoted to obtain constraints on model parameters from
the relic density, direct and indirect search of DM. In section-VIII, we lay the
conclusions with some outlook.

\section{The model for muon $g-2$ anomaly, neutrino mass and dark matter}

We consider the gauge extension of the SM with extra $U(1)_{L_\mu - L_\tau}$ symmetry
(from now on referred to as ``gauged $U(1)_{L_\mu - L_\tau}$ model'') where difference
between muon and tau lepton numbers is defined as a local gauge
symmetry~\cite{Baek:2001kca,Ma:2001md,Heeck:2011wj,Heeck:2010ub,
Heeck:2010pg,Ota:2006xr,Rodejohann:2005ru,Xing:2015fdg,Rivera-Agudelo:2016kjj,Choubey:2004hn,He:1991qd,Elahi:2015vzh,Araki:2015mya,Fuyuto:2015gmk,Heeck:2016xkh,Altmannshofer:2014cfa,Altmannshofer:2016oaq,
Crivellin:2015lwa,Crivellin:2015mga,Heeck:2014qea,Shuve:2014doa,Salvioni:2009jp,Yin:2009mc,Bi:2009uj,Adhikary:2006rf,Ma:2001tb,Bell:2000vh,He:1994aq,Kim:2015fpa,Harigaya:2013twa,Fuki:2006xw}.
The advantage of considering the gauged $U(1)_{L_\mu - L_\tau}$ model is that the theory
is free from any gauge anomaly without introduction of additional fermions. We break the
gauge symmetry $U(1)_{L_\mu - L_\tau}$ to a residual discrete symmetry $Z_2$ and explore
the possibility of having non-zero neutrino mass and a viable candidate of DM.

\subsection{Spontaneous symmetry breaking}
The spontaneous symmetry breaking of gauged $U(1)_{L_\mu - L_\tau}$ model is given by:
\begin{align}
\mathcal{G}^{}_{L_\mu - L_\tau} \mathop{\longrightarrow}^{\langle S \rangle,\langle \eta \rangle } \mathcal{G}_{SM} \times Z_2 
\mathop{\longrightarrow}^{\langle H \rangle }  SU(3)_C \times U(1)_{\rm em}\times Z_2\, ,
\end{align}
where 
\begin{align}
 &\mathcal{G}^{}_{L_\mu - L_\tau}\equiv SU(3)_C \times SU(2)_L \times U(1)_Y \times U(1)_{\Lmu}\, ,\nonumber \\
 &\mathcal{G}_{SM} \equiv SU(3)_C \times SU(2)_L \times U(1)_Y \nonumber 
\end{align}

At first, the spontaneous symmetry breaking of $\mathcal{G}^{}_{L_\mu - L_\tau}  \to \,
\mathcal{G}_{SM} \times Z_2$ is achieved by assigning non-zero vacuum expectation values
(vevs) to complex scalar field $S$ and $\eta$. The subsequent stage of symmetry breaking
$\mathcal{G}_{SM}\times Z_2 \to SU(3)_C \times U(1)_{\rm em} \times Z_2$ is obtained with
the SM Higgs $H$ providing masses to known charged fermions. The complete spectrum of the
gauged $U(1)_{L_\mu - L_\tau}$ model in light of DM and neutrino mass is provided in
Table\,I  where the respective quantum numbers are  presented under $SU(3)_C \times
SU(2)_L \times U(1)_Y \times U(1)_{\Lmu}$.  To the usual quarks and leptons, we have
introduced additional neutral fermions $N_e, N_\mu, N_\tau$ for light neutrino mass
generation via seesaw mechanism and a vector like Dirac fermion $\chi$ for the candidate
of DM, being odd under the residual discrete symmetry $Z_2$. we note that except $\chi$
all other particles are even under the $Z_2$ symmetry.

\begin{table}[h!]
\label{tab:SM}
\begin{center}
\begin{tabular}{|c|c|c|c|c|c|}
	\hline
			& Field	& $ SU(3)_C \times SU(2)_L\times U(1)_Y$  & $L_\mu$    & $L_\tau$    & $L_\mu - L_\tau$  \\
	\hline
	\hline
	Quarks	& $Q_L \equiv(u, d)^T_L$			& $(\textbf{3},\textbf{2},~~1/6)$ & $0$        & $0$         & $0$   \\
			& $u_R$							& $(\textbf{3},\textbf{1},~~2/3)$ & $0$        & $0$         & $0$   \\
			& $d_R$							& $(\textbf{3},\textbf{1},-1/3)$ & $0$        & $0$         & $0$   \\
			\hline
	Leptons	& $L_e \equiv(\nu_e,e^-)^T_L$	 & $(\textbf{1},\textbf{2},~  -1/2)$ & $0$        & $0$         & $0$    \\
			& $L_\mu \equiv(\nu_\mu,\mu^-)^T_L$	 & $(\textbf{1},\textbf{2},~  -1/2)$ & $1$        & $0$         & $1$    \\
			& $L_\tau \equiv(\nu_\tau,\tau^-)^T_L$	 & $(\textbf{1},\textbf{2},~  -1/2)$ & $0$        & $1$         & $-1$    \\
			& $e_{R}$					& $(\textbf{1},\textbf{1},~ -1)$ & $0$        & $0$         & $0$    \\
			& $\mu_{R}$					& $(\textbf{1},\textbf{1},~ -1)$ & $1$        & $0$         & $1$    \\
			& $\tau_{R}$					& $(\textbf{1},\textbf{1},~ -1)$ & $0$        & $1$         & $-1$    \\
\hline
			& $N_{e}$					& $(\textbf{1},\textbf{1},~ 0)$ & $0$        & $0$         & $0$    \\
			& $N_{\mu}$					& $(\textbf{1},\textbf{1},~ 0)$ & $1$        & $0$         & $1$    \\
			& $N_{\tau}$					& $(\textbf{1},\textbf{1},~ 0)$ & $0$        & $1$         & $-1$    \\
			& $\chi$					& $(\textbf{1},\textbf{1},~ 0)$ & $\_$        & $\_$       & $1$    \\
	\hline
	Scalars	& $H$							& $(\textbf{1},\textbf{2},~ 1/2)$ & $\_$        & $\_$         & $0$   \\
		& $S$						&
                                                                  $(\textbf{1},\textbf{1},~   0)$ & $\_$        & $\_$         & $1$    \\  
                        &$\eta$                      &$(\textbf{1},\textbf{1},~0)$   & $\_$      &$\_$   & $2$ \\

	\hline
	\hline
\end{tabular}
\caption{Particle content of the minimal $U(1)_{\Lmu}$ gauge extension of the SM 
         and their transformation under the SM gauge group.}
\end{center}
\end{table}
\subsection{Interaction Lagrangian}
The complete interaction Lagrangian for the gauged $U(1)_{L_\mu - L_\tau}$ model is given
by
\begin{align}
\label{eq:TheModel}
	\mathcal{L}_{L_\mu-L_\tau} 
	&=     i \, \overline{N}_{e} \slashed{\partial} \,N_{e} 
	     + i \, \overline{N}_{\mu} \left( \slashed{\partial} + i\,g_{\mu \tau} \,Z_\mu^\prime \gamma^\mu \right)\,N_{\mu}	
	     + i \, \overline{N}_{\tau} \left( \slashed{\partial} - i\,g_{\mu \tau} \,Z_\mu^\prime \gamma^\mu \right)\,N_{\tau} 
	\nonumber \\
		&~~~ - g_{\mu \tau}   \big(\overline{\mu} \gamma^\mu \mu   + \overline{\nu_\mu} \gamma^\mu P_L \nu_\mu
		          -\overline{\tau} \gamma^\mu \tau -\overline{\nu_\tau} \gamma^\mu P_L \nu_\tau \big)Z_\mu^\prime
	\nonumber \\
	&~~~ -M_{ee} \overline{N^c_e} N_e - ( \lambda_{e\mu} S^\star \overline{N^c_e} N_\mu + h.c ) - ( \lambda_{e\tau} S \overline{N^c_e} N_\tau + h.c ) \nonumber \\
	&~~~ - ( \lambda_{\mu\mu} \eta^\star \overline{N^c_\mu} N_\mu + h.c ) - ( \lambda_{\tau\tau} \eta \overline{N^c_\tau} N_\tau + h.c )
		\nonumber \\
	&~~~ -\left( Y_{ee} \overline{L_e} \tilde{H} N_e + Y_{\mu \mu} \overline{L_\mu} \tilde{H} N_\mu
	      + Y_{\tau \tau} \overline{L_\tau} \tilde{H} N_\tau+ h.c \right) 
	\nonumber \\
      &~~~ + i\, \overline{\chi}  \left( \slashed{\partial} + i\, g_{\mu \tau} \,Z_\mu^\prime \gamma^\mu \right)\,\chi - M_\chi \overline{\chi} \chi -f_\chi \overline{\chi^c}\chi \eta^\star
	\nonumber \\
	&~~~ - \frac{1}{4} F_{Z^\prime}^{\mu \nu}F^{Z^\prime}_{\mu \nu} 
	     + \frac{\epsilon}{4} F_{Z^\prime}^{\mu \nu}\,F_{\mu \nu}
	\nonumber \\
	&~~~ + |\left( \partial_\mu + \,i\,g_{\mu \tau}\,Z'_\mu \right) S|^2
	  - \mu_S^2 S^\dagger S +\lambda_S (S^\dagger S)^2    + |\left( \partial_\mu + \,i\,2 g_{\mu \tau}\,Z'_\mu \right) \eta |^2 - \mu_\eta^2 \eta^\dagger \eta + \lambda_\eta ( \eta^\dagger \eta )^2    \nonumber \\
	  &~~~+ \lambda_{HS} (H^\dagger H) (S^\dagger S) + \lambda_{H\eta} ( H^\dagger H) ( \eta^\dagger \eta ) + \lambda_{\eta S} ( \eta^\dagger \eta ) (S^\dagger S) + \mu_{\eta S} S S \eta^\star 
		     + \mathcal{L}_\text{SM} 
	  \, ,
\end{align}
Here $\mathcal{L}_\text{SM}$ is the SM Lagrangian. We denote here $Z_\mu^{\prime}$ as the
new gauge boson for $U(1)_{L_\mu-L_\tau}$ and the corresponding field strength tensor as
$F_{\mu \nu}^{Z^\prime} = \partial_\mu Z_\nu^\prime - \partial_\nu Z_\mu^\prime$. The
gauge coupling corresponding to $U(1)_{L_\mu-L_\tau}$ is defined as $g_{\mu \tau} \equiv
g'$ (as mentioned in section \ref{sec:introduction}).

\subsection{Scalar masses and mixing} \label{scalar_mixing}
The scalar potential of the model is given by
\begin{align}
\mathcal{V}(H,S) & = - \mu_H^2 H^\dagger H +  \lambda_H(H^\dagger H)^2 -  \mu_\eta^2 \eta^\dagger \eta + \lambda_\eta ( \eta^\dagger \eta )^2 - \mu_S^2 S^\dagger S + \lambda_S(S^\dagger S)^2  \nonumber \\
& + \lambda_{SH} (H^\dagger H) (S^\dagger S) + \lambda_{H\eta} ( H^\dagger H) ( \eta^\dagger \eta ) + \lambda_{\eta S} ( \eta^\dagger \eta ) (S^\dagger S) + \mu_{\eta S} S S \eta^\star
\end{align}
where $H$ is the SM Higgs doublet and $S$, $\eta$ are the complex scalar singlets under
SM, while charged under $U(1)_{L_\mu-L_\tau}$. The neutral complex scalars $H^0$ , $S$ and
$\eta$ can be parametrised as follows:
\begin{align}
& H^0  =\frac{1}{\sqrt{2} }(v_H+h)+  \frac{i}{\sqrt{2} } G^0\, , \nonumber \\
& S  = \frac{1}{\sqrt{2} }(v_S+s)+  \frac{i}{\sqrt{2} } A\,, \nonumber \\
&\eta = \frac{1}{\sqrt{2}} (v_\eta + \eta ) + \frac{i}{\sqrt{2} } B\,, \nonumber \\
\end{align}
The mass matrix for the neutral scalars is given by
\begin{equation}
\mathcal{M}^2 = 
\begin{pmatrix}
2 \lambda_H \, v_H^2     &  \lambda_{SH} \, v_H \, v_S   & \lambda_{H \eta} \, v_H \, v_\eta \\
\lambda_{SH}\, v_H\, v_S  &  2 \lambda_S\, v_S^2      & \lambda_{\eta S}\, v_S \, v_\eta + \mu_{\eta S}\, v_S  \\
\lambda_{H \eta} \, v_H \, v_\eta  &  \lambda_{\eta S}\, v_S \, v_\eta + \mu_{\eta S}\, v_S  & 2 \lambda_\eta \, v_\eta^2
\end{pmatrix}
\end{equation}
This is a symmetric mass matrix. So it can be diagonalised by a unitary matrix:
\begin{equation}
V^\dagger \mathcal{M}^2 V = \rm Diagonal(M_h^2, \, M_S^2,  \, M_\eta^2) 
\end{equation}
We identify $M_h$ as the physical mass of the SM Higgs, while $M_S$ and $M_\eta$ are the
masses of additional scalars $S$ and $\eta$ respectively. Since $S$ and $\eta$ are
singlets, their masses can vary from sub-GeV to TeV region. For a typical set of values:
$v_H=174 \rm GeV, v_S = 1200 \rm GeV, v_\eta=50 \rm GeV,
\lambda_H=0.2585,\lambda_{SH}=0.0005,\lambda_S=0.4, \lambda_\eta=10^{-5}, \lambda_{\eta
H}= 10^{-3},\lambda_{\eta S}=0.0015,\mu_{\eta S}= 0.1 \rm GeV$ , the physical masses are
found to be $M_h=125$ GeV, $M_S=1073$ GeV, $M_\eta=0.1$ GeV and the mixing between $h$ and
$\eta$ field is $\sin \theta_{\eta h}=8.7\times 10^{-7}$. We will study the importance of
$\eta$ field while calculating the relic abundance of DM. The mixing between $\eta$ and
$h$ field is required to be small as it plays a dominant role in the direct detection of
DM. We will show in Fig.\ref{fig:dd_relic} that if the mixing angle is large then it will
kill almost all the relic abundance parameter space. 

\subsection{Mixing in the Gauge Sector}
\begin{figure}[thb]
$$
\includegraphics[width=0.4\textwidth,natwidth=610,natheight=642]{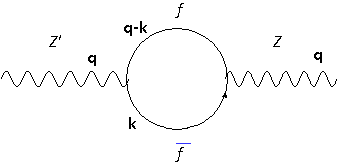}
$$
\caption{The mixing between the SM gauge boson $Z$ and the $U(1)_{L_\mu - L_\tau}$ gauge
boson $Z'$ arising through the exchange of muon and tauon families of leptons.}
\label{gaugemix}
\end{figure}

The breaking of gauged $U(1)_{L_\mu - L_\tau}$ symmetry by the vev of $S$ and $\eta$ gives
rise to a massive neutral gauge boson $Z'$ which couples to only muon and tauon families
of leptons. In the tree level there is no mixing between the SM gauge boson $Z$ and $Z'$.
However at one loop level, there is a mixing between $Z$ and $Z'$ through the exchange of
muon and tauon families of leptons as shown in the figure \ref{gaugemix}.
The loop factor can be estimated as
\begin{equation}
\Pi^{\mu \nu}(q^2) =- \left( q^2 g^{\mu \nu} - q^\mu q^\nu \right )
\frac{1}{3} \frac{1}{16 \pi^2} \left( g_{\mu \tau} \frac{ C_V g}{2 \cos
  \theta_W}\right ) \rm Log \left( \frac{m_f^2} {\Lambda^2} \right ) 
\end{equation}
where $\theta_W$ is the Weinberg angle, $C_V$ is the vector coupling of SM fermions with $Z$ boson, $\Lambda$ is the cut off scale of the theory and
$\text m_\text f$ is the mass of the charged fermion running in the loop. In the gauge basis, the mass
matrix is given by
\begin{equation}\label{gauge_boson_mass_matrix}
\mathcal{M}_2^2 =
\begin{pmatrix}
  M_{Z_0}^2    & \Pi \\
  \Pi      & \tilde{M}_{Z^\prime}^2
\end{pmatrix}
\end{equation}
where $\Pi$ is given by $\Pi = \Pi^{\mu \nu} * g_{\mu \nu}$ and $M_{Z_0}= 91.1876$ GeV.
Thus the mixing angle is given by
\begin{equation}
\tan 2 \theta_Z = \frac{2 \Pi} {\tilde{M}_{Z^\prime}^2 - M_{Z_0}^2}
\end{equation}
Diagonalising the mass matrix (\ref{gauge_boson_mass_matrix}) we get the eigen values:
\begin{eqnarray}\label {eq: mass}
M^2_Z = \frac{ M^2_{Z^0} - M^2_{Z^\prime} \sin^2 \theta_Z } {\cos ^2
  \theta_Z}  \nonumber \\ 
M^2_{Z^ \prime} = \frac{  \tilde{M}^2_{Z^\prime} - M_Z^2 \sin^2 \theta_Z } {\cos ^2 \theta_Z}
\end{eqnarray}
where $M_Z$ and $M_{Z^\prime}$ are the physical masses of $Z$ and $Z^\prime$ gauge bosons.
The mixing angle $\theta_Z$ has to chosen in such a way that the physical mass of Z-boson
should be obtained within the current uncertainty of the SM $Z$ boson mass~\cite{pdg}.  It
can be computed from equation (\ref{eq: mass}) as follows:
\begin{equation}
\frac {M_Z - M_{Z^0}} {M_{Z^0}} = \frac{M^2_{Z^0}-M^2_{Z^\prime}} {2
  M^2_{Z^0}} \tan^2 \theta_Z \leq 4.6 \times 10^{-5}
\end{equation}
For $M_{Z^\prime}-M_{Z^0} \gtrsim {M_{Z^0}}$ we get $\tan \theta_Z \lesssim 10 ^{-2}$.

\section{Muon $g-2$ Anomaly}\label{sec: anomaly}
The magnetic moment of muon is given by
\begin{equation}\label{anomaly}
\overrightarrow{\mu_\mu}= g_\mu \left (\frac{q}{2m} \right)
\overrightarrow{S}\,,
\end{equation}
where $g_\mu$ is the gyromagnetic ratio and its value is $2$ for a
structureless, spin $\frac{1}{2}$ particle of mass $m$ and charge
$q$. Any radiative correction,  which couples to the muon spin to the
virtual fields, contributes to its magnetic moment and is given by
\begin{equation}
a_\mu=\frac{1}{2} ( g_\mu - 2)
\end{equation}
At present there is a more than $3 \sigma$ discrepancy between the experimental
measurement \cite {Bennett:2006fi} and the SM prediction \cite {Miller:2007kk} of $a_\mu$
value. This is given by:
\begin{equation}
\Delta a_\mu = a_\mu ^{expt} - a_\mu ^{SM} = (295 \pm 88 ) \times 10^{-11}\,.
\end{equation}
In the present model, the new gauge boson $Z^\prime$ contributes to $\Delta a_\mu$ and is
given by~\cite{Baek:2008nz}

\begin{equation}
\Delta a_\mu = \frac{\alpha^\prime}{2\pi} \int_0^1  dx \frac{2m_\mu^2
  x^2(1-x)}{x^2 m_\mu^2 + (1-x) M_{Z^\prime}^2} \approx
\frac{\alpha^\prime}{2\pi} \frac {2m_\mu^2}{3 M_{Z^\prime}^2}\,,
\end{equation}
where $\alpha'=g_{\mu\tau}^2/4\pi$. The above equation implies that the discrepancy
between the experimental measurement \cite {Bennett:2006fi} and the SM prediction \cite
{Miller:2007kk} of $a_\mu$ value can be explained in a large region of parameter space as
shown in Fig. (\ref{fig:mz_copling}).

\section{Neutrino mass}\label{sec:neumass}
In order to account for tiny non-zero neutrino masses for light neutrinos, we extend the
minimal gauged $U(1)_{L_\mu - L_\tau}$ model with additional neutral fermions $N_e(0),
N_\mu(1), N_\tau(-1)$ where the quantum numbers in the parentheses are the $L_\mu -
L_\tau$ charge. The relevant Yukawa interaction terms are given by
\begin{align}
\mathcal{L}=&-\frac{1}{2} M_{ee} \overline{N^c_e} N_e -\frac{1}{2} M_{\mu \tau} \overline{N^c_\mu} N_\tau- ( \lambda_{e\mu} S^\star \overline{N^c_e} N_\mu + h.c )
             - ( \lambda_{e\tau} S \overline{N^c_e} N_\tau + h.c )
              \nonumber \\
         & -( \lambda_{\mu\mu} \eta^\star \overline{N^c_\mu} N_\mu +
           h.c ) - ( \lambda_{\tau\tau} \eta \overline{N^c_\tau}
           N_\tau + h.c ) \nonumber \\
           & -\left( Y_{ee} \overline{L_e} \tilde{H} N_e + Y_{\mu \mu} \overline{L_\mu} \tilde{H} N_\mu
             + Y_{\tau \tau} \overline{L_\tau} \tilde{H} N_\tau+ h.c \right) \nonumber \\
          =& -\frac{1}{2} N^T_\alpha \mathcal{C}^{-1} {M_R}_{\alpha \beta} N_\beta
             + {M_D}_{\alpha \beta} \overline{\nu_\alpha} N_\beta +\mbox{h.c.}
\end{align}
where the Dirac and Majorana neutrino mass matrices are given by
\begin{align}
M_R =\begin{pmatrix}
               M_{ee}      &  \lambda_{e\mu} v_ S
    & \lambda_{e\tau} v_ S \\
               \lambda_{e\mu} v_ S      &  \lambda_{\mu \mu}
v_ \eta     & M_{\mu \tau}  \\
               \lambda_{e\tau} v_ S     &  M_{\mu \tau}    &
               \lambda_{\tau \tau} v_ \eta   
               \end{pmatrix}\, , \quad
M_D =\begin{pmatrix}
               Y_{ee} v_ H      &  0    & 0  \\
               0     &  Y_{\mu \mu}v_H    & 0  \\
               0     &  0    & Y_{\tau \tau} v_H  
               \end{pmatrix}\, , \quad               
\end{align}
Using seesaw approximation, the light neutrino mass matrix can be read as
\begin{align}
m_\nu \simeq - M_D M^{-1}_R M^T_D \,.
\end{align}

We illustrate here a specific scenario where not only the resulting Dirac neutrino mass
matrix is diagonal but also degenerate. As a result, we can express $M_D = m_d \mathbb{I}_{3\times3}$.
One can express heavy Majorana neutrino mass matrix in terms of light neutrino mass matrix as
\begin{align}
M_R = m^2_d m^{-1}_{\nu}\, .
\end{align}
Thus, one can reconstruct $M_R$ using neutrino oscillation parameters and $m_d\simeq 10^{-4}~$GeV.
As we know that light neutrino mass matrix is diagonalised by the PMNS mixing matrix as
$$m^{\rm diag.}_\nu = U_{\rm PMNS}^\dagger m_\nu U_{\rm PMNS}^* = \mbox{diag.}\{m_1, m_2, m_2 \}$$
where $m_i$ are the light neutrino mass eigenvalues. The PMNS mixing matrix is generally parametrized
as
\begin{equation}
U_{\rm PMNS}=\left(\begin{array}{ccc}  c^{}_{12}
c^{}_{13} & s^{}_{12} c^{}_{13} & s^{}_{13} e^{-i\delta} \\
-s^{}_{12} c^{}_{23} - c^{}_{12} s^{}_{13} s^{}_{23} e^{i\delta} &
c^{}_{12} c^{}_{23} - s^{}_{12} s^{}_{13} s^{}_{23} e^{i\delta} &
c^{}_{13} s^{}_{23} \\ s^{}_{12} s^{}_{23} - c^{}_{12} s^{}_{13}
c^{}_{23} e^{i\delta} & -c^{}_{12} s^{}_{23} - s^{}_{12} s^{}_{13}
c^{}_{23} e^{i\delta} & c^{}_{13} c^{}_{23} \end{array} \right)\cdot P \;
\end{equation}
where $c^{}_{ij} \equiv \cos\theta^{}_{ij}$, $s^{}_{ij} \equiv \sin\theta^{}_{ij}$ (for $ij = 12, 13, 23$),
and $P ={\rm diag.}\{1, e^{i\alpha}, e^{i\beta} \}$. Here we denoted Dirac phase as $\delta$ and Majorana phases
as $\alpha, \beta$.

For a numerical example, we consider the best-fit values of the oscillation parameters,
the atmospheric mixing angle $\theta_a\equiv \theta_{23} \simeq 41.2^\circ$, solar angle
$\theta_s \equiv\theta_{12} \simeq 34.2^\circ$, the reactor mixing angle $\theta_r\equiv \theta_{13}
\simeq 9^\circ$, and the Dirac CP phase $\delta=0.8 \pi$ (Majorana phases assumed to be zero here
for simplicity i.e, $\alpha, \beta=0$). The PMNS mixing matrix for this best-fit oscillation parameters
is estimated to be
\begin{eqnarray} \label{eq:inputU}
U_{\rm PMNS} = \left( \begin{array}{ccc}
0.8168                & 0.5552               & -0.1265-0.0919\, i \\
-0.3461 - 0.0510\, i  & 0.6604 - 0.0347\, i  &  0.6634   \\
0.4551 - 0.0563\, i   & -0.5028 - 0.0382\, i &  0.7316
\end{array} \right)\, .
\end{eqnarray}

We also use the best-fit values of
mass squared differences $\Delta m^2_{\rm s} \equiv m^2_2-m^2_1 = 2.5 \times 10^{-5} \mbox{eV}^{2}$
and $\Delta m^2_{\rm {a}}\equiv |m^2_3-m^2_1| = 7.56 \times 10^{-3} \mbox{eV}^{2}$. As we do not know
the sign of $\Delta m^2_{\rm {a}}$, the pattern of light neutrinos could be normal hierarchy (NH)
with $m_1 < m_2 < m_3$,
      \begin{equation*}
         m_2 = \sqrt{m_1^2 +\Delta m^2_{\rm {s}}}\;,\qquad
         m_3 = \sqrt{m_1^2 +\Delta m^2_{\rm {a}}}\;, \label{eq1:NH_m2_m3}
      \end{equation*}
or, the inverted hierarchy (IH) with $m_3 < m_1 < m_2$,
      \begin{equation*}
         m_1 = \sqrt{m_3^2 +\Delta m^2_{\rm {a}}}\;,\qquad
         m_2 = \sqrt{m_3^2 +\Delta m^2_{\rm {a}}+\Delta m^2_{\rm {s}}}\;. \label{eq1:IH_m1_m2}\,.
      \end{equation*}
Now, one can use these oscillation parameters and $m_d \simeq 10^{-4}~$GeV, the mass matrix
for heavy neutrinos is expressed as
\begin{align}
M_R = 10^{-8}\,\mbox{GeV}^2 \left(U_{\rm PMNS} m^{\rm diag.}_\nu U_{\rm PMNS}^T \right)^{-1}\, .
\end{align}
Using $m_1=0.001~$eV, the masses for heavy neutrinos are found to be $M_{N_1}\simeq 100~$GeV,
$M_{N_2}\simeq 1000~$GeV and $M_{N_3}\simeq 8000~$GeV. The same algebra can be extended for inverted
hierarchy and quasi-degenerate pattern of light neutrinos for deriving structure of $M_R$.

\section{Relic abundance of dark matter}
The local $U(1)_{\Lmu}$ is broken to a remnant $Z_2$ symmetry under which $\chi$ is odd
and all other fields are even. As a result $\chi$ becomes a viable candidate for DM. We
explore the parameter space allowed by relic abundance and null detection of DM at direct
search experiments in the following two cases: \\
(a) in absence of $\eta$ \\
(b) in presence of $\eta$. \\

\subsection{Relic abundance in absence of $\eta$}
\begin{figure}
    \includegraphics[width=\textwidth,natwidth=610,natheight=642]{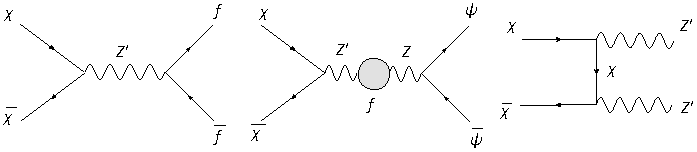}
     \caption{Possible annihilation channels for relic abundance of DM, where $f$ represents muon and 
tauon families of leptons while $\psi$ represents the SM fermion.}
     \label{fig:coan}
\end{figure}

\begin{figure}[h]
\centering
\includegraphics [width=0.5 \textwidth] {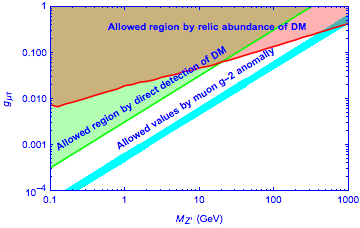}
\caption{Constraints from correct relic abundance of DM, muon $g-2$ anomaly and null detection of DM at LUX 
using $Z-Z^\prime$ mixing $10^{-2}$ are shown in the plane of $M_{Z^\prime}$ and $g_{\mu\tau}$. }
\label{fig:mz_copling}
\end{figure}

For simplicity, we assume that the right handed neutrinos $N_\mu$ and $N_\tau$ as well as
the scalar field $S$ are heavier than the $\chi$ mass. Now in absence of
$\eta$~\footnote{In absence of $\eta$ the neutrino mass will not be affected.}, the
relevant diagrams that contribute to the relic abundance of DM $\chi$ are shown in Fig.
(\ref{fig:coan}).  
Since the null detection of DM at direct search experiments, such as Xenon-100 and LUX
restricts the $Z-Z^\prime$ mixing to be small ($\tan \theta_z < 10^{-2}$), the dominant
contribution to relic abundance, below the threshold of $Z^\prime$, comes from the
s-channel annihilation: $\bar{\chi} \chi \to \bar{\psi} \psi, \bar{f} f$ through the
exchange of $Z^\prime$. Due to the resonance effect this cross-section dominates. We have
shown in Fig. \ref{fig:mz_copling}, the correct relic abundance of DM in the plane of
$M_{Z^\prime}$ and $g_{\mu\tau}$. Below the red line the annihilation cross-section
through $Z^\prime$ exchange is small due to small gauge coupling and therefore, we always
get an over abundance of DM. The constraints from muon $g-2$ anomaly and direct detection
of DM via $Z-Z^\prime$ mixing are also shown in the same plot for comparison.  From Fig.
\ref{fig:mz_copling}, we see that in a large parameter space we can not get any point
which satisfies both relic abundance of DM as well as muon $g-2$ anomaly constraints.
Therefore, it does not serve our purpose. We resolve the above mentioned issues in
presence of the scalar field $\eta (2)$, where the number inside the parenthesis is the
charge under $U(1)_{\Lmu}$.    

\subsection{Relic abundance in presence of $\eta$} 
\begin{figure}[h]
\centering
\includegraphics [width=0.4 \textwidth] {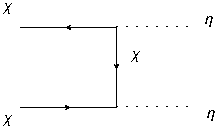}
\caption{Dominant annihilation channel for relic abundance of DM in the region of small $g_{\mu\tau}$.}
\label{fig:feyann}
\end{figure}
\begin{figure}[h]
\centering
\includegraphics [width =0.5 \textwidth] {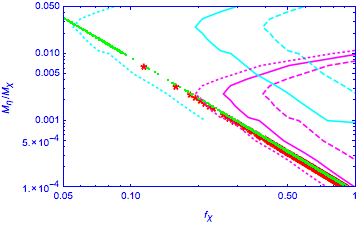}
\caption{Constraints on the parameter space satisfying correct relic abundance (shown by Green and Red points) and null detection of dark 
matter at LUX (shown by Magenta and Cyan lines). For $M_\eta =0.1$ GeV (Magenta lines) we have used $\sin \theta_{\eta h}=5 \times 10^{-8}$ 
(dashed-line), $7.5 \times 10^{-8}$ (solid-line) and $1 \times 10^{-7}$ (dotte-line), while for $M_\eta = 1$ GeV (Cyan lines), we have 
used $\sin \theta_{\eta h}=5 \times 10 ^{-6}$ (dashed-line), $1  \times 10^{-5}$ (solid-line) and $3.5 \times 10 ^ {-5}$ (dotted-line).}
\label{fig:dd_relic}
\end{figure}

In presence of the SM singlet scalar field $\eta (2)$, the new annihilation channels
$\bar{\chi}\chi \to \eta^\dagger \eta$, shown in Fig. \ref{fig:feyann}, and $\bar{\chi}
\chi \rightarrow h \eta$ open up in addition to the earlier mentioned channels, shown in
fig. (\ref{fig:coan}). However, in the region of small gauge coupling $g_{\mu \tau}$, the
dominant channel for the relic abundance of DM is $\bar{\chi}\chi \to \eta^\dagger \eta$.
The other channel: $\bar{\chi} \chi \rightarrow h \eta$ is suppressed due to the small
mixing angle $\sin \theta_{\eta h}$, required by null detection of DM at direct search
experiments. We assume that the mass of $\eta$ to be less than a GeV as discussed in
section \ref{scalar_mixing}. In this case the analytic expression for the cross-section of
$\bar{\chi}\chi \to \eta^\dagger \eta$ is given by: 
\begin{equation}\label{chichi-etaeta}
\langle \sigma | v \rangle (\chi \chi \rightarrow \eta \eta)  = \frac{1}{128 \pi } \frac{1}{\left(1-\frac{M_\eta^2}{2M_\chi^2}\right)^2}
\frac{f_\chi^4}{M_\chi^2} \, \left( 1- \frac{M_\eta^2}{M_\chi^2}
\right)^{3/2} 
\end{equation}
From the above expression we observe that the cross-section goes as
$\frac{f_\chi^4}{M_\chi^2}$ for $M_\eta\ll M_\chi$.  Fixing $M_\eta \sim 0.1$ GeV and
varying the DM mass $M_\chi$ and the coupling $f_\chi$, we have shown in Fig.
\ref{fig:dd_relic} the allowed region in the plane of $M_\eta/M_\chi$ and $f_\chi$ for the
correct relic abundance. The green points show the value of analytic approximation
\ref{chichi-etaeta}, while the red points reveal the result from full calculation using
\texttt{micrOMEGAs}~\cite{micromegas}.  The matching of both points indicates that the
contribution to relic abundance is solely coming from the $\bar{\chi}\chi \to \eta^\dagger
\eta$ channel. From the Fig. \ref{fig:dd_relic} it is clear that as the ratio
$\frac{M_\eta}{M_\chi}$ decreases, {\it i.e.,} $M_\chi$ increases for a fixed value of
$M_\eta$, we need a large coupling to get the correct relic abundance. For comparison, we
also show the DM-nucleon spin independent elastic cross-section: $nn\to \chi \chi$
mediated through the $\eta-h$ mixing, in the same plot. We see that the allowed mixing
angle by LUX data is quite small. 

\section{Direct Detection}
We constrain the model parameters from null detection of DM at direct search experiments
such as Xenon-100~\cite{Aprile:2012nq} and LUX~\cite{Akerib:2013tjd} in the following two
cases: \\
a. In the absence  of $\eta$\\
b. In the presence of $\eta$ .\\
We show that in absence of $\eta$ field, the elastic scattering of DM with nucleon through
$Z-Z^\prime$ mixing give stringent constraint on the model parameters: $M_{Z^\prime}$ and
$g_{\mu \tau}$, as depicted in fig.  (\ref{fig:mz_copling}). On the other hand, in the
presence of $\eta$ field , the elastic scattering will be possible through the $\eta - h$
mixing, while inelastic scattering with nucleon will be possible via $Z-Z^\prime$ mixing.
In the following we discuss in details the possible constraints on model parameters. 

\subsection{Direct Detection in absence of $\eta$}
\begin{figure}[h!]
\centering
\includegraphics[width=0.2\textwidth,natwidth=610,natheight=642]{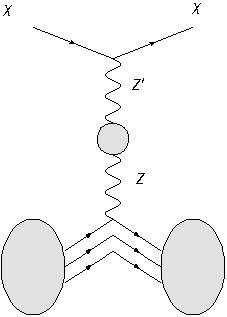}
\caption{Feynman diagrams for direct detection of DM through scattering with nuclei via the exchange of $Z-Z^\prime$ mixing.}
\label{feyn:DD-Zmediated}
\end{figure}

While the direct detection of DM through its elastic scattering with nuclei is a very
challenging task, the splendid current sensitivity of present direct DM detection
experiments might allow to set stringent limits on parameters of the model, or hopefully
enable the observation of signals in near future. In absence of $\eta$ field, the elastic
scattering between singlet fermion DM with nuclei is displayed in
Fig.\,\ref{feyn:DD-Zmediated}.

The spin independent DM-nucleon cross-section mediated via the loop induced $Z-Z^\prime $
mixing is given by~\cite{Goodman:1984dc,Essig:2007az}
\begin{equation}\label{DM-nucleon-Z}
\sigma_{\rm SI}^Z = \frac{1}{64 \pi A^2 }\mu_r^2 \tan^2 \theta_Z
\frac{G_F}{2\sqrt{2}} \frac{g_{\mu \tau}^2}{M_{Z^\prime}^2} \left[Z
  \frac{f_p}{f_n} + (A-Z) \right]^2 f_n^2   \, ,
\end{equation}
where $A$ is the mass number of the target nucleus, $\mu_r=M_\chi m_n/(M_\chi + m_n)
\approx m_n$ is the reduced mass, $m_n$ is the mass of nucleon (proton or neutron) and
$f_p$ and $f_n$ are the interaction strengths (including hadronic uncertainties) of DM
with proton and neutron respectively. Here $Z$ is the atomic number of the target nucleus.

For simplicity we assume conservation of isospin, {\it i.e.} $f_p/f_n=1$. The value of
$f_n$ is varied within the range: $0.14 < f_n < 0.66$ \cite{hadronic_uncertainty}. If we take $f_n \simeq 1/3$, the
central value, then from Eqn. (\ref{DM-nucleon-Z}) we get the total cross-section per
nucleon to be
\begin{equation}
\sigma_{\rm SI}^Z \simeq 7.6 \times 10^{-46} {\rm cm}^2 \tan^2 \theta_Z \frac{g_{\mu\tau}^2}{M_{Z^\prime}^2}  \,.
\end{equation}
for the DM mass of $33$ GeV.

Since the $Z$-boson mass puts a stringent constraint on the mixing parameter $\tan
\theta_Z$ to be $\mathcal{O} ( 10^{-2}-10^{-4})$~\cite{Hook:2010tw,Babu:1997st}, we choose
the maximum allowed value ($10^{-2}$) and plot the smallest spin independent direct DM
detection cross-section ($7.6\times 10^{-46} cm^2$), allowed by LUX, in the plane of
$g_{\mu \tau}$ versus $M_{Z^\prime}$ as shown in fig (\ref{fig:mz_copling}). The plot
follows a straight line as expected from equation (\ref{DM-nucleon-Z}) and shown by the
green line in fig. (\ref{fig:mz_copling}). Any values above that line will be allowed by
the LUX limit.

\subsection{Direct Detection in presence of $\eta$}
In presence of the $\eta$ field both elastic and inelastic scattering between DM and the
nuclei is possible.  The elastic scattering is mediated through $\eta - h$ mixing while
inelastic scattering is mediated by the $Z-Z^\prime$ mixing.
\begin{figure}[h!]
\centering
\includegraphics [width=0.2\textwidth]{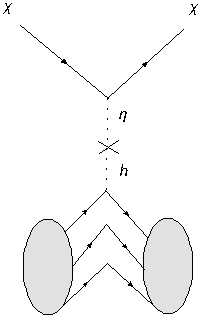}
\caption{Elastic scattering of DM with nuclei through $\eta - h$ mixing.}
\label{fig: Scalar_dd}
\end{figure}
\subsubsection{Elastic scattering of dark matter}

The spin-independent scattering of DM with nuclei is a t-channel exchange diagram as shown
in Fig.\,\ref{fig: Scalar_dd} through the mixing of scalar singlet $\eta $ with the SM
Higgs $H$.  The elastic scattering cross section $\sigma^{n}_{SI}$ off a nucleon is given
by\cite{Goodman:1984dc,Essig:2007az} :
\begin{equation}\label{dd_scalar_mixing}
\sigma_{\rm SI}^{\eta h}= \frac{\mu_r^2}{\pi A^2} \left[ Z f_p + (A-Z) f_n\right]^2
\end{equation}
where $\mu_r$ is the reduced mass, Z and A are respectively atomic and mass number of the
target nucleus. In the above equation $f_p$ and $f_n$ are the effective interaction
strengths of the DM with the proton and neutron of the target nucleus and are given by: 
\begin{equation}\label{effective_strength}
f_{p,n}=\sum_{q=u,d,s} f_{T_q}^{p,n} \alpha_q \frac{m_{p,n}}{m_q} + \frac{2}{27} f_{TG}^{p,n} \sum_{q=c,t,b}\alpha_q \frac{m_{p,n}}{m_q}
\end{equation}
with 
\begin{equation}\label{alpha_value}
\alpha_q=\frac{f_\chi \sin \theta_{\eta h}}{M_\eta^2} \left( \frac{m_q}{v_H}\right)
\end{equation}
In Eq. (\ref{effective_strength}), the different coupling strengths between the DM and
light quarks are given by~\cite{DM_review} $f^{(p)}_{Tu}=0.020\pm 0.004$,
$f^{(p)}_{Td}=0.026\pm 0.005$,$f^{(p)}_{Ts}=0.118\pm 0.062$, $f^{(n)}_{Tu}=0.014\pm
0.004$,$f^{(n)}_{Td}=0.036\pm 0.008$,$f^{(n)}_{Ts}=0.118\pm 0.062$. The coupling of DM
with the gluons in target nuclei is parameterised by 
\begin{equation}\label{Gluon-interaction}
f^n_{TG}=1-\sum_{q=u,,d,s}f^n_{Tq}\,. 
\end{equation}
Thus from Eqs. \ref{dd_scalar_mixing}, \ref{effective_strength}, \ref{alpha_value},
\ref{Gluon-interaction}, the spin-independent DM-nucleon interaction through $\eta-h$
mixing is given by:
\begin{eqnarray}\label{ddscalar}
\sigma_{SI}^{\eta h} &=& \frac{\mu_r^2 f_\chi^2 \sin^2 \theta_{\eta h} }{\pi A^2 M_\eta^4 }\nonumber\\
&\times & \left[Z\frac{m_p}{v_H} \left(f_{T_u}^{p}+ f_{T_d}^{p}+f_{T_s}^{p}+\frac{2}{9}f_{TG}^{p}\right)+ (A-Z)\frac{m_n}{v_H}  \left(f_{T_u}^{n}+ f_{T_d}^{n}+f_{T_s}^{n}+\frac{2}{9}f_{TG}^{n}\right) \right]^2 
\end{eqnarray}
In the above equation, the only unknowns are $f_\chi$, $\sin \theta_{\eta h}$ and
$M_\eta$. So using the current limit on spin-independent scattering cross-section from
Xenon-100~\cite{Aprile:2012nq} and LUX~\cite{Akerib:2013tjd} one can constrain these
parameters $f_\chi$ and $M_\eta$ for a fixed value of mixing angle $\sin \theta_{\eta h}$.
Here we use LUX bound and the corresponding contour lines are drawn in the fig.
\ref{fig:dd_relic} by choosing $M_\eta =0.1 {\rm Gev}$ (magenta lines) for three values of
mixing angles: $\sin\theta_{\eta h} = 5\times 10^{-8}$(dashed), $\sin\theta_{\eta h}=7.5
\times 10^{-8}$(solid) and $\sin \theta _{\eta h} = 10^{-7}$ (dotted). Similarly for
another value of  $M_\eta=1 \rm GeV$ (cyan lines), we have drawn three lines for
$\sin\theta_{\eta h} = 5 \times 10^{-6}$(dashed), $\sin\theta_{\eta h}= 10^{-5}$ (solid)
and $\sin \theta _{\eta h} = 3.5 \times 10^{-5}$ (dotted). The regions on the right of the
respective lines are excluded by LUX data. From fig. (\ref{fig:dd_relic}), we see that for
a constant value of $M_\eta$, if $\sin \theta_{\eta h}$ decreases then the curves shift
towards higher value of $f_\chi$.

\subsubsection{Inelastic scattering of dark matter}
As we discussed above inelastic scattering \cite{TuckerSmith:2001hy} of the DM with the target nuclei is possible
via $Z-Z^\prime$ mixing. Let us rewrite the DM Lagrangian in presence of $\eta$ field as\cite{Cui:2009xq,Arina:2011cu,Arina:2012fb,Arina:2012aj} :
\begin{eqnarray}
\mathcal{L}_{DM} &=& i\, \overline{\chi}  \left( \slashed{\partial} + i\, g_{\mu \tau} \,Z_\mu^\prime \gamma^\mu \right)\,\chi \nonumber\\ 
&& - M_\chi \overline{\chi} \chi  -\frac{1}{2} f_1\left(\overline{\chi^C} P_L \chi +\rm h.c \right) \eta^\star -
\frac{1}{2} f_2 \left ( \overline{\chi^C} P_R \chi +\rm h.c \right) \eta^\star \,,
\end{eqnarray}

where $f_1$ and $f_2$ are the interaction strengths to left and right components of the
vector-like fermion $\chi$. When $\eta$ gets a vev, the DM gets  small Majorana mass
$m_L=f_1 v_\eta$ and $m_R=f_2 v_\eta$. The presence of small Majorana mass terms for the
DM split the Dirac state into two real Majorana states $\chi_1$ and $\chi_2$. The
Lagrangian in terms of the new eigenstates is given as 
\begin{equation}\label{eq: dm}
\begin{split}
\mathcal{L}_{DM} = \, &\frac{1}{2} \overline{\chi_1} i \gamma^\mu
\partial_\mu \chi_1 -\frac{1}{2} M_1 \overline{\chi_1} \chi_1 
 + \frac{1}{2} \overline{\chi_2} i \gamma^\mu
\partial_\mu \chi_2-\frac{1}{2} M_2 \overline{\chi_2} \chi_2 
+ ig_{\mu \tau} \overline{\chi_2} \gamma^\mu \chi_1  \; Z_\mu^\prime    \\
& +\frac{1}{2} g_{\mu \tau} \;  \frac{m_-}{M_\chi}
\left(\overline{\chi_2}\gamma^\mu \gamma^5 \chi_2 - \overline{\chi_1}
  \gamma^\mu \gamma^5 \chi_1 \right ) Z_\mu^\prime+ \mathcal{O}
(\frac{m_-^2}{M_\chi^2})\\
& + \frac{1}{2} \left( f_1 \cos^2\theta - f_2 \sin^2 \theta \right)
\overline{\chi_1}\chi_1 \eta + \frac{1}{2} \left( f_2 \cos^2\theta - f_1 \sin^2 \theta \right)
\overline{\chi_2}\chi_2 \eta \, ,
  \end{split}
\end{equation}
where $\sin\theta$ is the mixing angle , $M_1$ and $M_2$ are the two mass eigenvalues and
are given by 
\begin{eqnarray}
M_1= M_\chi-m_+ , M_2 = M_\chi +m_+ \\
m_\pm= \frac{ m_L \pm m_R }{2}
\end{eqnarray}
From the above expression the dominant gauge interaction is
off-diagonal, and the diagonal interaction is suppressed as $\frac{m_-}{M_\chi} \ll 1$.
The mass splitting between the two mass eigen states is given by:
\begin{equation}
\delta = M_2-M_1 = 2 m_+=  (f_1+f_2) v_\eta\,.
\end{equation}
\begin{figure}[h]
\centering
\includegraphics[width = 0.2 \textwidth] {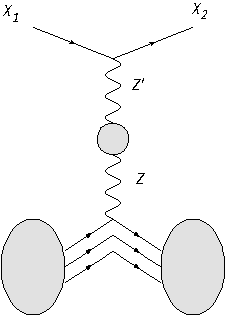}
\caption{Inelastic scattering of DM with the target nucleus through the $Z-Z^\prime$ mixing.}
\label{fig: inelas}
\end{figure}
The inelastic scattering with the target nucleus due to $Z-Z^\prime$ mixing is shown in
Fig. \ref{fig: inelas}. The occurrence of this process solely depends on the mass
splitting between the two states. In fact, the minimum velocity of the DM needed to
register a recoil inside the detector is given by \cite{TuckerSmith:2001hy,Cui:2009xq,Arina:2011cu,Arina:2012fb,Arina:2012aj} :
\begin{equation}
v_{\rm min}= c\sqrt{\frac{1}{2 m_n E_R}} \left( \frac{m_n E_R}{\mu_r} + \delta \right)\,,
\end{equation}    
where $E_R$ is the recoil energy of the nucleon. If the mass splitting is above a few
hundred keV, then it will be difficult to excite $\chi_2$. So the inelastic scattering
will be forbidden.

\section{Indirect detection}

We now look at the compatibility of the present framework with indirect detection signals
of DM and in particular AMS-02 positron data. Recently, the AMS-02 experiment reported the
results of high precision measurement of the cosmic ray positron fraction in the energy
range of $0.5-500$~GeV\cite{Aguilar:2013qda,Accardo:2014lma}. This result further
confirmed the measurement of an excess in the positron fraction above $10~$GeV as observed
by PAMELA\cite{Adriani:2008zr,Adriani:2010ib} and FERMI-LAT\cite{FermiLAT:2011ab}.  The
usual explanation for this excess is through DM annihilation producing the required flux
of positrons.  However such an excess was not observed in the antiproton flux by
PAMELA\cite{pamela-pbar}, thus suggesting a preference for leptonic annihilation channels.
Recently AMS-02 also announced results from their measurement of the antiproton flux,
which suggests a slight excess above 100 GeV\cite{ams2pbar}.  But this was found to be
within error of the modelling of secondary astrophysical production\cite{cirelli}.  In
this context we consider the $L_\mu-L_\tau$ symmetry where the DM dominantly annihilates
to muons which then subsequently decay to produce electrons.  This ensures a softer
distribution of positrons thereby providing a better fit to the experimental data.

For theoretical explanation for AMS-02 positron excess through DM annihilations in the
$L_\mu-L_\tau$ symmetric extension of SM we have to calculate propagation of cosmic rays
in the galaxy.  In order to do this calculation, the propagation of cosmic rays is treated
as a diffusion process and one therefore solves the appropriate diffusion equation.  Here
we calculate the flux of the cosmic ray electrons (primary and secondary) as well as
secondary positrons at the position of the sun after propagating through the galaxy.  The
propagation equation for charged cosmic rays is given by\cite{strong}
\begin{align}\nonumber
 \frac{\partial\psi(\vec{r},p)}{\partial t}=q&+\vec{\nabla}\cdot
 \left(D_{xx}\vec\nabla\psi-\vec{V_c}\psi\right)+\frac{\partial}{\partial p}p^2
 D_{pp}\frac{\partial}{\partial p}\frac{1}{p^2}\psi-\frac{\partial}{\partial p}
 \left[\dot p\psi-\frac{p}{3}\left(\vec\nabla\cdot\vec{V_c}\right)\psi\right]\\
 &-\frac{1}{\tau_f}\psi-\frac{1}{\tau_r}\psi\label{diff}
\end{align}
where $\psi$ is the cosmic ray density, $\dot{p}$ gives the energy loss of cosmic rays,
$D_{xx(pp)}$ is the diffusion coefficient in spatial (momentum) coordinates while the last
two terms represent the fragmentation and radioactive decay of cosmic ray nuclei.  The
diffusion coefficient is parameterized as $D_{xx}=D_{0xx}E^{-\gamma_e-\delta}$.  The
primary spectrum of cosmic ray electrons is modeled by
\begin{equation} \psi = \frac{N}{2}\frac{L}{D_{0xx}}E^{-\gamma_e-\delta}
\end{equation}
where N is a normalization constant and $L$ is the half height of the cylindrical
diffusion zone.  The parameters for propagation of cosmic rays are $D_0$, $\delta$, $N$,
$L$, $v_a$ (Alfven velocity), $V_c$ and $\gamma_e$.  We use the GALPROP package
\cite{galprop} to solve the diffusion equation in Eq.~\ref{diff} using a diffusive
re-acceleration model of diffusion.  The cosmic ray primary and secondary electron flux as
well as the secondary positron flux which constitute the astrophysical background are thus
obtained.  The positron flux from DM annihilations is calculated using
\texttt{micrOMEGAs}\cite{micromegas} while the gauged $U(1)_{L_\mu - L_\tau}$ model is
implemented in \texttt{micrOMEGAs} with the help of LanHEP\cite{semenov}.  The ratio of
the DM positron signal thus obtained, to the total astrophysical background gives the
positron fraction.

\begin{figure}[ht!]
	\centering
	\includegraphics{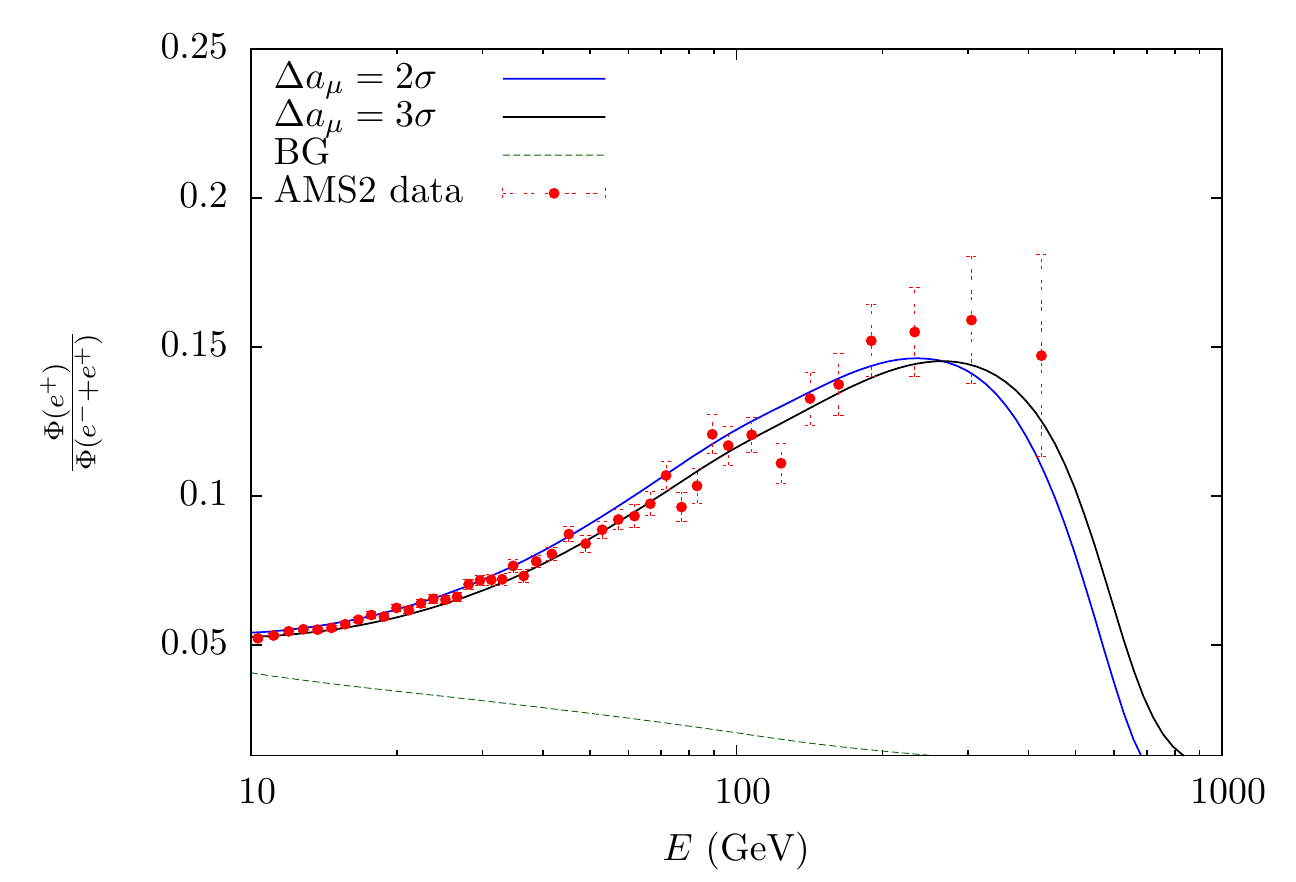}
	\caption{Ratio of positron flux to the total $(e^- +e^+)$ flux against 
	         energy of the cosmic rays with AMS-02(2014) data\cite{Accardo:2014lma}
		 for benchmark points listed in Table~\ref{bmp}.  The blue curve is for
		 the benchmark point with $\Delta a_\mu=2\sigma$ while the black curve is
	 	for $\Delta a_\mu=3\sigma$.}
	\label{ams2}
\end{figure}

The key feature of the model is that the DM does not couple to quarks at tree
level and hence we do not see any observable contribution to the antiproton flux.  We
therefore only plot the positron fraction against DM energy in Fig.~\ref{ams2}, for two
benchmark points chosen such that they satisfy the relic density constraint from
PLANCK\cite{PLANCK} and the contribution to $\Delta a_\mu$ is $2\sigma$ and $3\sigma$
respectively.  The parameters for the two chosen benchmark points are listed in
Table~\ref{bmp}.
We find that for the best fit to AMS-02 data in the current scenario requires $M_\chi\gtrsim
500$ GeV.  Also for satisfying the relic density constraint we need $M_{Z^\prime}\sim 500$
GeV for $M_\chi\gtrsim 500$ GeV.  

\begin{table}
	\centering
	\begin{tabular}{|c|c|c|c|c|c|}
	\hline
	$M_\chi$ (GeV) & $M_{Z^\prime}$ (GeV) & $g_{\mu \tau}$ & $\Omega h^2$ & $\Delta
	a_\mu$ & Boost factor\\\hline
	710 & 838 & 0.35 & 0.116 & $2\sigma$ & 720 \\
	800 & 782 & 0.4 & 0.113 & $3\sigma$ & 800 \\\hline
	\end{tabular}
	\caption{Benchmark point which satisfies relic density and fits the AMS2 positron
	fraction data\cite{Accardo:2014lma}.}
	\label{bmp}
\end{table}


\section{Conclusion}
We discussed a gauged $U(1)_{L_\mu-L_\tau}$ extension of the SM in light of the non-zero
neutrino mass, DM and the observed muon $g-2$ anomaly which is a more than $3\sigma$
discrepancy between the experimental measurement and the SM prediction. In adition to that,
three right handed neutrinos $N_e, N_\mu, N_\tau$ and a Dirac fermion $\chi$ were
introduced which are charged under $U(1)_{L_\mu-L_\tau}$ symmetry except $N_e$ which is a complete singlet fermion. The
$U(1)_{L_\mu-L_\tau}$ was allowed to break to a survival $Z_2$ symmetry at a TeV scale by
giving vev to a SM singlet scalar $S$ which bears an unit $L_\mu-L_\tau$ charge. The vev
of $S$ gave masses not only to the additional gauge boson $Z^\prime$, but also to the
right handed neutrinos: $N_e, N_\mu, N_\tau$. As a result, below electroweak symmetry
breaking, the light neutrinos acquired masses through the type-I seesaw mechanism. Under
the survival $Z_2$ symmetry, $\chi$ was chosen to be odd while rest of the particles were
even. Thus $\chi$ became an excellent candidate of DM.

We obtained the relic abundance of DM via its annihilation to muon and tauon families of
leptons through the exchange of $Z^\prime$ gauge boson. It is found that for $Z^\prime$
mass greater than 100 MeV and its coupling to leptons: $g_{\mu \tau}> 5 \times 10^{-3}$
correct relic abundance can be obtained. On the other hand, the muon $g-2$ anomaly
required smaller values of the gauge coupling $g_{\mu \tau}$ for $Z^\prime$ mass greater
than 100 MeV (see fig. \ref{fig:mz_copling}).  So the two problems could not be solved
simultaneously. Therefore, we introduced an additional scalar $\eta$ which is doubly
charged under $U(1)_{L_\mu-L_\tau}$ but singlet under the SM gauge group. In presence of
$\eta$, the DM dominantly annihilates to $\eta$ fields. As a result we found a large
region of parameter space in which the constraints from muon $g-2$ anomaly and relic
abundance of DM could be satisfied simultaneously. 

The hitherto null detection of DM at direct search experiments, such as LUX, also give
strong constraints on the model parameters as discussed in Fig. \ref{fig:mz_copling}. We
found that for $Z^\prime$ mass greater than 100 MeV we need the corresponding gauge
coupling: $g_{\mu \tau} > 2 \times 10^{-4}$. In fact such values of $g_{\mu \tau}$ hardly
agree with muon $g-2$ constraint. However, in presence of $\eta$ we can allow small
values of the gauge coupling $g_{\mu \tau}$ which can satisfy muon $g-2$ anomaly while
direct detection limit can be satisfied through $\eta-h$ mixing diagram. 

The annihilation of DM to only muon and tauon families of leptons dictates its nature to
be leptophilic. So, the observed positron flux by PAMELA, Fermi-LAT and recently by AMS-02
in the cosmic ray shower with suppressed anti-proton flux could be explained in our model.
We showed in Fig. \ref{ams2} that the constraints from muon $g-2$ anomaly and AMS-02
positron excess can be satisfied simultaneously in our model.

\section*{Aknowledgement}
The work of SR is supported by the University of Adelaide and the Australian Research Council through the ARC Center of Excellence in Particle Physics at the Terascale.
 Narendra Sahu is partially supported by the Department of Science and Technology, Govt. of India under the financial Grant SR/FTP/PS-209/2011.


\begin{thebibliography}{99}

\bibitem{Miller:2007kk} 
  J.~P.~Miller, E.~de Rafael and B.~L.~Roberts,
  Rept.\ Prog.\ Phys.\  {\bf 70}, 795 (2007)
  [hep-ph/0703049].

\bibitem{Fukuda:2001nk} 
  S.~Fukuda {\it et al.} [Super-Kamiokande Collaboration],
  Phys.\ Rev.\ Lett.\  {\bf 86}, 5656 (2001)
  [hep-ex/0103033].

\bibitem{DM_review} 
G. Bertone, D. Hooper and J. Silk, 
Phys. Rept. 405, 279 (2005), arXiv:hep-ph/0404175; 
G. Jungman, M. Kamionkowski and K. Griest, 
Phys. Rept. 267, 195 (1996), arXiv:hep-ph/9506380.


\bibitem{wmap} G.~Hinshaw {\it et al.} [WMAP Collaboration],
  Astrophys.\ J.\ Suppl.\  {\bf 208}, 19 (2013)
  [arXiv:1212.5226 [astro-ph.CO]].
\bibitem{PLANCK} 
P.~A.~R.~Ade {\it et al.}  [Planck Collaboration],
Astron.\ Astrophys.\  {\bf 571}, A16 (2014), arXiv:1303.5076 [astro-ph.CO].


  
  

\bibitem{Baek:2001kca} 
  S.~Baek, N.~G.~Deshpande, X.~G.~He and P.~Ko,
  Phys.\ Rev.\ D {\bf 64}, 055006 (2001)
  [hep-ph/0104141].


\bibitem{Ma:2001md} 
  E.~Ma, D.~P.~Roy and S.~Roy,
  Phys.\ Lett.\ B {\bf 525}, 101 (2002)
  [hep-ph/0110146].


\bibitem{Heeck:2011wj} 
  J.~Heeck and W.~Rodejohann,
  Phys.\ Rev.\ D {\bf 84}, 075007 (2011)
  [arXiv:1107.5238 [hep-ph]].

\bibitem{Heeck:2010ub} 
  J.~Heeck and W.~Rodejohann,
  AIP Conf.\ Proc.\  {\bf 1382}, 144 (2011)
  [arXiv:1012.2298 [hep-ph]].


\bibitem{Heeck:2010pg} 
  J.~Heeck and W.~Rodejohann,
  J.\ Phys.\ G {\bf 38}, 085005 (2011)
  [arXiv:1007.2655 [hep-ph]].


\bibitem{Ota:2006xr} 
  T.~Ota and W.~Rodejohann,
  Phys.\ Lett.\ B {\bf 639}, 322 (2006)
  [hep-ph/0605231].


\bibitem{Rodejohann:2005ru} 
  W.~Rodejohann and M.~A.~Schmidt,
  Phys.\ Atom.\ Nucl.\  {\bf 69}, 1833 (2006)
  [hep-ph/0507300].


\bibitem{Xing:2015fdg} 
  Z.~z.~Xing and Z.~h.~Zhao,
  arXiv:1512.04207 [hep-ph].


\bibitem{Rivera-Agudelo:2016kjj} 
  D.~C.~Rivera-Agudelo and A.~Pérez-Lorenzana,
  arXiv:1603.02336 [hep-ph].
  
\bibitem{Choubey:2004hn} 
  S.~Choubey and W.~Rodejohann,
  Eur.\ Phys.\ J.\ C {\bf 40}, 259 (2005)
  [hep-ph/0411190].


\bibitem{He:1991qd} 
  X.~G.~He, G.~C.~Joshi, H.~Lew and R.~R.~Volkas,
  Phys.\ Rev.\ D {\bf 44}, 2118 (1991).
\bibitem{Altmannshofer:2016oaq} 
  W.~Altmannshofer, M.~Carena and A.~Crivellin,
  arXiv:1604.08221 [hep-ph].
\bibitem{Altmannshofer:2014cfa} 
  W.~Altmannshofer, S.~Gori, M.~Pospelov and I.~Yavin,
  Phys.\ Rev.\ D {\bf 89}, 095033 (2014)
  [arXiv:1403.1269 [hep-ph]].
\bibitem{Heeck:2016xkh} 
  J.~Heeck,
  Phys.\ Lett.\ B {\bf 758}, 101 (2016)
  [arXiv:1602.03810 [hep-ph]].
\bibitem{Fuyuto:2015gmk} 
  K.~Fuyuto, W.~S.~Hou and M.~Kohda,
  Phys.\ Rev.\ D {\bf 93}, no. 5, 054021 (2016)
  [arXiv:1512.09026 [hep-ph]].

\bibitem{Araki:2015mya} 
  T.~Araki, F.~Kaneko, T.~Ota, J.~Sato and T.~Shimomura,
  Phys.\ Rev.\ D {\bf 93}, no. 1, 013014 (2016)
  [arXiv:1508.07471 [hep-ph]].
\bibitem{Crivellin:2015lwa} 
  A.~Crivellin, G.~D'Ambrosio and J.~Heeck,
  Phys.\ Rev.\ D {\bf 91}, no. 7, 075006 (2015)
  [arXiv:1503.03477 [hep-ph]].
\bibitem{Crivellin:2015mga} 
  A.~Crivellin, G.~D'Ambrosio and J.~Heeck,
  Phys.\ Rev.\ Lett.\  {\bf 114}, 151801 (2015)
  [arXiv:1501.00993 [hep-ph]].
\bibitem{Heeck:2014qea} 
  J.~Heeck, M.~Holthausen, W.~Rodejohann and Y.~Shimizu,
  Nucl.\ Phys.\ B {\bf 896}, 281 (2015)
  [arXiv:1412.3671 [hep-ph]].
\bibitem{Shuve:2014doa} 
  B.~Shuve and I.~Yavin,
  Phys.\ Rev.\ D {\bf 89}, no. 11, 113004 (2014)
  [arXiv:1403.2727 [hep-ph]].
\bibitem{Salvioni:2009jp} 
  E.~Salvioni, A.~Strumia, G.~Villadoro and F.~Zwirner,
  JHEP {\bf 1003}, 010 (2010)
  [arXiv:0911.1450 [hep-ph]].
\bibitem{Yin:2009mc} 
  P.~f.~Yin, J.~Liu and S.~h.~Zhu,
  Phys.\ Lett.\ B {\bf 679}, 362 (2009)
  [arXiv:0904.4644 [hep-ph]].
\bibitem{Bi:2009uj} 
  X.~J.~Bi, X.~G.~He and Q.~Yuan,
  Phys.\ Lett.\ B {\bf 678}, 168 (2009)
  [arXiv:0903.0122 [hep-ph]].
\bibitem{Adhikary:2006rf} 
  B.~Adhikary,
  Phys.\ Rev.\ D {\bf 74}, 033002 (2006)
  [hep-ph/0604009].

\bibitem{Ma:2001tb} 
  E.~Ma and D.~P.~Roy,
  hep-ph/0111385.
\bibitem{Bell:2000vh} 
  N.~F.~Bell and R.~R.~Volkas,
  Phys.\ Rev.\ D {\bf 63}, 013006 (2001)
  [hep-ph/0008177].
\bibitem{He:1994aq} 
  X.~g.~He,
  hep-ph/9409237.
\bibitem{Kim:2015fpa} 
  J.~C.~Park, S.~C.~Park and J.~Kim,
  Phys.\ Lett.\ B {\bf 752}, 59 (2016)
  [arXiv:1505.04620 [hep-ph]].
\bibitem{Harigaya:2013twa} 
  K.~Harigaya, T.~Igari, M.~M.~Nojiri, M.~Takeuchi and K.~Tobe,
  JHEP {\bf 1403}, 105 (2014)
  [arXiv:1311.0870 [hep-ph]].
\bibitem{Fuki:2006xw} 
  K.~Fuki and M.~Yasue,
  Nucl.\ Phys.\ B {\bf 783}, 31 (2007)
  [hep-ph/0608042].




\bibitem{Elahi:2015vzh} 
  F.~Elahi and A.~Martin,
  Phys.\ Rev.\ D {\bf 93}, no. 1, 015022 (2016)
  [arXiv:1511.04107 [hep-ph]].

\bibitem{Carena:2004xs} 
  M.~Carena, A.~Daleo, B.~A.~Dobrescu and T.~M.~P.~Tait,
  Phys.\ Rev.\ D {\bf 70}, 093009 (2004)
  [hep-ph/0408098].

  
\bibitem{Aprile:2012nq} 
  E.~Aprile {\it et al.} [XENON100 Collaboration],
  Phys.\ Rev.\ Lett.\  {\bf 109}, 181301 (2012)
  [arXiv:1207.5988 [astro-ph.CO]].
\bibitem{Akerib:2013tjd} 
  D.~S.~Akerib {\it et al.} [LUX Collaboration],
  Phys.\ Rev.\ Lett.\  {\bf 112}, 091303 (2014)
  [arXiv:1310.8214 [astro-ph.CO]].
  
  


\bibitem{Adriani:2008zr} 
  O.~Adriani {\it et al.} [PAMELA Collaboration],
  Nature {\bf 458}, 607 (2009)
  [arXiv:0810.4995 [astro-ph]].
\bibitem{Adriani:2010ib} 
  O.~Adriani {\it et al.},
  Astropart.\ Phys.\  {\bf 34}, 1 (2010)
  [arXiv:1001.3522 [astro-ph.HE]].


\bibitem{FermiLAT:2011ab} 
  M.~Ackermann {\it et al.} [Fermi-LAT Collaboration],
  Phys.\ Rev.\ Lett.\  {\bf 108}, 011103 (2012)
  [arXiv:1109.0521 [astro-ph.HE]].
  
\bibitem{Aguilar:2013qda} 
  M.~Aguilar {\it et al.} [AMS Collaboration],
  Phys.\ Rev.\ Lett.\  {\bf 110}, 141102 (2013).
 

  
\bibitem{Accardo:2014lma} 
  L.~Accardo {\it et al.} [AMS Collaboration],
  Phys.\ Rev.\ Lett.\  {\bf 113}, 121101 (2014).

 \bibitem{pdg} 
 K.A. Olive et al. (Particle Data Group), Chin. Phys. C, 38, 090001 (2014).




\bibitem{Bennett:2006fi} 
  G.~W.~Bennett {\it et al.} [Muon g-2 Collaboration],
  Phys.\ Rev.\ D {\bf 73}, 072003 (2006)
  [hep-ex/0602035].

\bibitem{Baek:2008nz} 
  S.~Baek and P.~Ko,
  JCAP {\bf 0910}, 011 (2009)
  [arXiv:0811.1646 [hep-ph]].




\bibitem{micromegas}
	G.~B\'elanger, F.~Boudjema, A.~Pukhov and A.~Semenov,
	  Comput.\ Phys.\ Commun.\  {\bf 192}, 322 (2015)
      [arXiv:1407.6129 [hep-ph]].



\bibitem{Goodman:1984dc} 
  M.~W.~Goodman and E.~Witten,
  Phys.\ Rev.\ D {\bf 31}, 3059 (1985).
 








\bibitem{Essig:2007az} 
  R.~Essig,
  Phys.\ Rev.\ D {\bf 78}, 015004 (2008)
  [arXiv:0710.1668 [hep-ph]].
\bibitem{hadronic_uncertainty} R. Koch, Z.Physik C 15 161 (1982) ;
J. Gasser, H. Leutwyler and M. E. Sainio, Phys. Lett. B 253 260 (1991) ;
M. M. Pavan, R. A. Arndt, I. I. Strakovski and R. L. Workman, PiN Newslett. 16 110 (2002) ;
A. Bottino, F. Donato, N. Fornengo and S. Scopel, Phys. ReV. D 78 083520 (2008), [arXiv:0806.4099] .


\bibitem{Hook:2010tw} 
  A.~Hook, E.~Izaguirre and J.~G.~Wacker,
  Adv.\ High Energy Phys.\  {\bf 2011}, 859762 (2011)
  [arXiv:1006.0973 [hep-ph]].

\bibitem{Babu:1997st} 
  K.~S.~Babu, C.~F.~Kolda and J.~March-Russell,
  Phys.\ Rev.\ D {\bf 57}, 6788 (1998)
  [hep-ph/9710441].

\bibitem{TuckerSmith:2001hy} 
  D.~Tucker-Smith and N.~Weiner,
  Phys.\ Rev.\ D {\bf 64}, 043502 (2001)
  [hep-ph/0101138].

 
\bibitem{Cui:2009xq} 
  Y.~Cui, D.~E.~Morrissey, D.~Poland and L.~Randall,
  JHEP {\bf 0905}, 076 (2009)
  [arXiv:0901.0557 [hep-ph]].

\bibitem{Arina:2011cu} 
  C.~Arina and N.~Sahu,
  Nucl.\ Phys.\ B {\bf 854}, 666 (2012)
  [arXiv:1108.3967 [hep-ph]].
\bibitem{Arina:2012fb} 
  C.~Arina, J.~O.~Gong and N.~Sahu,
  Nucl.\ Phys.\ B {\bf 865}, 430 (2012)
  [arXiv:1206.0009 [hep-ph]].

\bibitem{Arina:2012aj} 
  C.~Arina, R.~N.~Mohapatra and N.~Sahu,
  Phys.\ Lett.\ B {\bf 720}, 130 (2013)
  [arXiv:1211.0435 [hep-ph]].


 





\bibitem{pamela-pbar}
	 O.~Adriani {\it et al.} [PAMELA Collaboration],
	   Phys.\ Rev.\ Lett.\  {\bf 105}, 121101 (2010)
	       [arXiv:1007.0821 [astro-ph.HE]].
\bibitem{ams2pbar}
	AMS-02 Collaboration, Talks at the ’AMS Days at CERN’, 15-17 April, 2015.
\bibitem{cirelli}
	G.~Giesen, M.~Boudaud, Y.~G\'enolini, V.~Poulin, M.~Cirelli, P.~Salati and
	P.~D.~Serpico,
  JCAP {\bf 1509}, no. 09, 023 (2015)
      [arXiv:1504.04276 [astro-ph.HE]].
\bibitem{strong}
      A.~W.~Strong, I.~V.~Moskalenko and V.~S.~Ptuskin,
        Ann.\ Rev.\ Nucl.\ Part.\ Sci.\  {\bf 57}, 285
	(2007)
	    [astro-ph/0701517].
\bibitem{galprop}
	I.~V.~Moskalenko and A.~W.~Strong,
	  Astrophys.\ J.\  {\bf 493}, 694 (1998)
	      [astro-ph/9710124];
    A.~E.~Vladimirov {\it et al.},
      Comput.\ Phys.\ Commun.\  {\bf 182}, 1156
      (2011)
	  [arXiv:1008.3642 [astro-ph.HE]].


\bibitem{semenov} 
	  A.~V.~Semenov,
	    hep-ph/9608488.

      







\end{thebibliography}
\end{document}